\documentclass[prd,aps,showpacs]{revtex4}
\usepackage{amsmath,amssymb}
\usepackage{epsf}
\newcommand{\be}{\begin{equation}}
\newcommand{\ee}{\end{equation}}
\newcommand{\bea}{\begin{eqnarray}}
\newcommand{\eea}{\end{eqnarray}}

\begin{document}

\title{Strings in Yang-Mills-Higgs theory coupled to gravity}
\author{L.G.~Aldrovandi}

\affiliation{Departamento\ de F\'{\i}sica, Universidad Nacional de
La Plata,
      C.C. 67, (1900) La Plata, Argentina.
      }
\date{\today}

\begin{abstract}

Non-Abelian strings for an Einstein-Yang-Mills-Higgs theory are
explicitly constructed. We consider $N_f$ Higgs fields in the
fundamental representation of the $U(1)\times SU(N_c)$ gauge group
in order to have a color-flavor $SU(N_c)$ group remaining
unbroken. Choosing a suitable ansatz for the metric,
Bogomol'nyi-like first order equations are found and rotationally
symmetric solutions are proposed. In the $N_f = N_c$ case,
solutions are local strings and are shown to be truly non-Abelian
by parameterizing them in terms of orientational collective
coordinates. When $N_f > N_c$, the solutions correspond to
semilocal strings which, beside the orientational degrees of
freedom, acquire additional collective coordinates parameterizing
their transverse size. The low-energy effective theories for the
correspondent moduli are found, showing that all zero modes are
normalizable in presence of gravity, even in the semilocal case.

\end{abstract}
\maketitle

\section{Introduction}

It is well known that solitons (kinks, vortices, monopoles) play a
central role in field theories, both at classical and quantum
levels. In particular, a new type of (string-like) vortex
solutions, called non-Abelian strings, was found quite recently in
certain supersymmetric \cite{Hanany:2003hp}-\cite{Hanany:2004ea}
and non-supersymmetric \cite{Gorsky:2004ad} gauge theories. More
in detail, these strings arise in certain $U(N_c)$
Yang-Mills-Higgs theories with $N_f (\geq N_c)$ flavors and are
mainly characterized by the presence of collective coordinates
related to the orientation of the flux-tube in the internal
color-flavor space.  Due to these orientational moduli space, the
above-mentioned strings behave as genuinely non-Abelian, leading
to a number of new exciting phenomena: from confinement in ${\cal
N}=1$ SQCD \cite{Shifman:2007kd}-\cite{Eto:2007hf} and
field-theoretic prototypes of D branes/strings
\cite{Shifman:2003uh},\cite{Sakai:2005sp} to applications in
cosmology as cosmic strings
\cite{Hashimoto:2005hi},\cite{Eto:2006db}. Concerning the topic of
cosmic strings (for a review and references see
\cite{2}-\cite{1}), non-Abelian strings were originally introduced
in this context as candidates to realize a mechanism proposed by
Polchinski \cite{1} through which gauge solitons could mimic the
reconnection properties of fundamental strings.

The gravitational properties of vortex-like configurations were
extensively studied in the past \cite{2}. In a field-theoretic
context, the simplest and most common model in which these
configurations appear is the Einstein-Maxwell-Higgs model
\cite{Garfinkle:1985hr}-\cite{Comtet:1987wi}. In the general case,
Einstein-Maxwell-Higgs theories support two types of string
solutions. These can be distinguished by their asymptotic
geometries, which must be one of the two Levi-Civita metrics,
\be
    ds^2 = dt^2 - (dx^3)^2- d\rho^2 - (a_1 \rho + a_2) d\theta^2,
    \label{cono}
\ee
whence the cone, or
\be
    ds^2 = (b_1 \rho + b_2)^{\frac 4 3} (dt^2 - (dx^3)^2) - d\rho^2 - (b_1 \rho + b_2)^{-\frac 2 3} d\theta^2,
    \label{nocono}
\ee
which is a Kasner metric. The Kasner branch does not have the
required characteristics to describe a `standard'  cosmic string,
and thus it is usually disregarded in  physical applications. Each
of the metrics (\ref{cono}) and (\ref{nocono}) has two totally
different behaviors. The behavior depends on the strength of the
gravitational coupling, which is measured by the parameter $G\xi$
(where $\xi$ is the symmetry-breaking scale). That is, for
$G\xi<<1$, which includes the GUT symmetry-breaking scale and most
of the applications in cosmology, $a_1$ and $b_1$ are positives
and then, (\ref{cono}) is the standard cone
\cite{Garfinkle:1985hr} and (\ref{nocono}) is the asymptotic form
of Melvin's magnetic universe \cite{Christensen:1999wb}. For
supermassive strings, which have $G\xi\gtrsim 1$, $a_1$ and $b_1$
are negatives and then there is a conical singularity both in the
Kasner-type metric \cite{Laguna:1989rx} and in the conical metric
\cite{Ortiz:1990tn}. When the Bogomol'nyi limit
\cite{Bogomolny:1975de} of the Einstein-Maxwell-Higgs theory is
considered, the critical coupling yields to a considerable
simplification of the problem, since all second-order equations
can be replaced by a curved-space analogue of the Bogomol'nyi
equations \cite{Linet:1987qu}. String-like solutions in the
Bogomol'nyi limit of the theory were analyzed in
\cite{Linet:1987qu},\cite{Comtet:1987wi} when only one Higgs field
is present. The generalization to more than one Higgs field were
considered in \cite{Gibbons:1992gt}, where the solutions obtained
correspond to gravitating semilocal vortices. An immediate
consequence of the Bogomol'nyi equations is that the metric
necessarily takes the conical asymptotic form (\ref{cono}) (see
\cite{Linet:1987qu},\cite{Ortiz:1990tn}). This feature is also
inherited by the supergravity extensions of the Maxwell-Higgs
model, as was shown very recently in the embedding of Abelian
cosmic strings into ${\cal N}=1$ and ${\cal N}=2$ supergravity
\cite{Dvali:2003zh}.

On the other hand, analogous investigations on string solutions in
Einstein-Yang-Mills-Higgs theories have remained, for some reason,
rather incomplete (for a recent analysis of Einstein-Yang-Mills
strings see \cite{Gal'tsov:2006tp}). It is the purpose of this
work to fill this gap looking for Einstein-Yang-Mills-Higgs
theories which support local and semilocal non-Abelian string
solutions. Since we are interested in the Bogomol'nyi limit of the
model, only string metrics asymptotically conical will be
analyzed. After presenting in section II the four dimensional
Einstein-Yang-Mills-Higgs model with gauge group $U(1)\times
SU(N_c)$, $N_f$ flavors and an arbitrary Higgs potential, we
consider in section III a  suitable ansatz for the metric which
reduces the equations of motion to a set of first order
(Bogomol'nyi) equations for a certain quartic potential. Next, in
section IV we find gravitating local non-Abelian strings by
considering a rotationally symmetric ansatz in the model with
$N_c=N_f$. These strings are shown to have a non-Abelian character
due to the existence of a set of orientational collective
coordinates. In section V we proceed  to do a similar analysis in
the $N_c<N_f$ case, yielding to gravitating semilocal non-Abelian
strings. Apart from the orientational degrees of freedom,
semilocal strings acquire new collective coordinates related to
variations of the transverse size. In this  case, we are able to
find, in the large transverse size limit, explicit analytic
solutions, not only for the matter fields, but also for the
space-time metric. The section ends showing that in the limit of a
very large transverse size of the string, semilocal solutions
approximate two-dimensional sigma-model instantons on $
Gr(N_c,N_f)$ (i.e., Grassmannian lumps). In section VI we use the
Manton procedure to obtain the low-energy effective theory of the
moduli for both the local and semilocal cases. In the case of
local strings, we have only orientational moduli and we find that
the correspondent effective theory is a two dimensional ${\bf
CP}^{N_c-1}$ sigma-model, just the same as in flat space-time. For
semilocal strings, where there are not only orientational modes
but also size moduli, we find in the large transverse size limit
three different effective theories depending on which the value of
the parameter $G\xi$ is. Quite remarkably, in contrast to what
happens in flat space-time, all moduli become normalizable in the
gravitating string case. Finally in section VII we present a
summary and a discussion of our results.

\section{The Theory}

The field content of the theory is given by a space-time metric
$g_{\mu\nu}$ where $\mu,\nu,...=0,1,2,3$ are space-time indices, a
$SU(N_c)\times U(1)$ gauge field ${\rm A}_\mu $ and $N_f$ complex
scalars $\phi$. As well as the $SU(N_c)\times U(1)$ gauge
symmetry, the Lagrangian also enjoys a $SU(N_f)$ flavor symmetry.
Under these two groups, the scalar fields transform as
$(\bf{N_c,\bar N_f})$. Thus, $\phi$ can be seen as an
$N_c\!\times\! N_f$ matrix $\phi=\phi^a_{\;r}$, where the indices
$a,b,...=1,2,...,N_c$ refer to the gauge group and
$r,s,...=1,2,...,N_f$ to the flavor group.

We represent the gauge fields in terms of matrices in the
fundamental representation of $SU(N_c)\times U(1)$, that is, ${\rm
A_\mu} = A_\mu^A T^A + i/\sqrt{2N_c} A_\mu$, where $T^A$
($A,B,...=1,2,..., N_c^2-1$) are the generators of the ${\bf N_c}$
representation de $SU(N_c)$. We use anti-Hermitian generators
$T^A$ satisfying
\begin{equation}
   [(T^A)_{\; b}^a]^*=-(T^A)_{\; a}^b, \;\;\;\;\;\;\;\;\;\;\;\;\;
   [T^A,T^B]=f^{ABC}T^C, \;\;\;\;\;\;\;\;\;\;\;\;\; (T^AT^B)_{\;
   a}^a=-\frac 1 2 \delta^{AB}.
\end{equation}

The action takes the form
\begin{equation}
    {\cal S}=  \int d^4 x \sqrt{g}\left\{ -\frac{1}{16\pi G}R
    - \frac{1}{4e_1^2}g^{\mu\rho}g^{\nu\sigma}F_{\mu\nu}F_{\rho\sigma}
    - \frac{1}{4e_2^2}g^{\mu\rho}g^{\nu\sigma}F^A_{\mu\nu}F^A_{\rho\sigma}
    + {\cal D}_\mu \bar\phi^r{\cal D}^\mu \phi_r  - V(\phi,\bar\phi)\right\}
    \label{action}
\end{equation}
where
\begin{equation}
    \bar\phi^r_{\;a}\equiv(\phi^a_{\;r})^*,\;\;\;\;\;\;\;
    F_{\mu\nu}=\partial_\mu A_\nu-\partial_\nu A_\mu,\;\;\;\;\;\;\;
    F^A_{\mu\nu}=\partial_\mu A_\nu^A-\partial_\nu A_\mu^A
    -f^{ABC}A_\mu^B A_\nu^C
\end{equation}
and
\begin{equation}
    {\cal D}_\mu \phi^a_{\;r} \equiv \partial_\mu \phi^a_{\;r}
    - \frac{i}{\sqrt{2N_c}}A_\mu \phi^a_{\;r}- A_\mu^A (T^A)_{\;b}^a \phi^b_{\;r},\;\;\;\;\;\;\;
    {\cal D}_\mu \bar\phi^r_{\;a} \equiv \partial_\mu \bar\phi^r_{\;a}
    + \frac{i}{\sqrt{2N_c}}A_\mu \bar\phi^r_{\;a}+ A_\mu^A (T^A)_{\;a}^b \bar\phi^r_{\;b}.
\end{equation}
Also, we use the conventions
$R^\mu_{\;\nu\rho\sigma}=\Gamma^\mu_{\nu\sigma,\rho}
-\Gamma^\mu_{\nu\rho,\sigma}+\Gamma^\mu_{\omega\sigma}\Gamma^\omega_{\nu\rho}
-\Gamma^\mu_{\omega\rho}\Gamma^\omega_{\nu\sigma}$,
$R_{\mu\nu}=R^\rho_{\;\mu\rho\nu}$, signature $(+,-,-,-)$ and $g =
-{\rm det}g_{\mu\nu}$.

For simplicity we do not write  gauge group indices which are
summed, e.g.,
\begin{equation}
   (\bar\phi^rT^A\phi_r)\equiv(T^A)_{\;b}^a\bar\phi^r_{\;a}\phi^b_{\;r}
\end{equation}

The equations of motion obtained from the variation of the action
are
\begin{eqnarray}
    && G_{\mu\nu}=R_{\mu\nu} - \frac 1 2 g_{\mu\nu} R = 8\pi G
    (T_{\mu\nu}^{\rm U(1)}+T_{\mu\nu}^{\rm SU(N_c)}+T_{\mu\nu}^{\rm mat}),\label{eins} \\
    && \partial_\mu (\sqrt g g^{\mu\rho} g^{\nu\sigma}F_{\rho\sigma})=
    e_1^2 \sqrt g j^{\nu}\vphantom{\frac 1 2},\label{yang}\\
    && (\partial_\mu \delta^{AB}+ f^{ABC} A_\mu^C)(\sqrt g g^{\mu\rho}
    g^{\nu\sigma}F^B_{\rho\sigma})= e_2^2 \sqrt g j^{A\nu},\vphantom{\frac 1 2}\label{gany}\\
    && [{\cal D}_\mu (\sqrt g g^{\mu\nu}{\cal D}_\mu \phi)]^a_{\;r} =
    -\sqrt g \frac{\partial V}{\partial \bar\phi^r_{\; a}},\label{higgs}
\end{eqnarray}
where the stress-energy tensors and the gauge currents are given
by
\begin{eqnarray}
   T_{\mu\nu}^{\rm U(1)} &=& \frac{1}{e_1^2}\left(-F_{\mu\rho}F_{\nu}^{\;\rho} +
   \frac 1 4 g_{\mu\nu} F_{\rho\sigma}F^{\rho\sigma}\right),\\
   T_{\mu\nu}^{\rm SU(N_c)} &=& \frac{1}{e_2^2}\left(-F^A_{\mu\rho}F_{\nu}^{A\,\rho} +
   \frac 1 4 g_{\mu\nu} F^A_{\rho\sigma}F^{A\rho\sigma}\right),\\
   T_{\mu\nu}^{\rm mat} &=&  {\cal D}_\mu \bar\phi^r{\cal D}_\nu \phi_r
   +{\cal D}_\nu \bar\phi^r{\cal D}_\mu \phi_r  - g_{\mu\nu} {\cal D}_\rho \bar\phi^r
   {\cal D}^\rho \phi_r + g_{\mu\nu} V,\vphantom{\frac 1 2}\\
   j^\mu &=& \frac {i}{\sqrt{2N_c}}({\cal D}^\mu\bar\phi^r\phi_r-\bar\phi^r {\cal D}^\mu\phi_{r}),\\
   j^{A\mu} &=& {\cal D}^\mu\bar\phi^rT^A\phi_{r}-\bar\phi^r T^A{\cal D}^\mu\phi_{r}\vphantom{\frac 1 2}.
\end{eqnarray}

In order to study cosmic string solutions, we assume the metric,
gauge and matter fields to be static and symmetric under $x^3$
translations. We also restrict to purely magnetic configurations.
Thus, we will consider the following ansatz for the metric and
gauge fields,
\begin{eqnarray}
    && ds^2 = L^2 dt^2 - K^2 (dx^3)^2+ h_{ij}dx^idx^j, \\
    && {\rm A}_\mu dx^\mu = {\rm A}_idx^i,\quad\quad\quad\quad i,j=1,2
\end{eqnarray}
where the fields $L$, $K$, ${\rm A}_i$ and $h_{ij}$ depend only on
the two transverse coordinates $x^k$ ($k=1,2$).

With this ansatz the components of the Ricci tensor take the form
\begin{eqnarray}
    R_{00} &=& -L[(h^{ij}L_{,j})_{,i} + \gamma^k_{ik}h^{ij}L_{,j}] - \frac L K h^{ij}
    K_{,i} L_{,j} = - \frac L K (KL^{,i})_{;i},\\
    R_{33} &=& K[(h^{ij}K_{,j})_{,i} + \gamma^k_{ik}h^{ij}K_{,j}] + \frac K L h^{ij}
    L_{,i} K_{,j} =  \frac K L (LK^{,i})_{;i},\\
    R_{ij} &=& r_{ij} - \frac 1 L (L_{,i,j} - \gamma^k_{ij} L_{,k} ) - \frac 1 K (K_{,i,j}
    - \gamma^k_{ij} K_{,k} )\\
    &=& r_{ij} - \frac {L_{,i;j}} {L} - \frac {K_{,i;j}} {K},\\
    R_{i0} &=& R_{i3} \;=\; 0,\quad R_{03} \;=\; 0,\vphantom{\frac 1 2}
\end{eqnarray}
where $\gamma^k_{ij}$, $r_{ij}$ and ``;" denote the connection,
Ricci tensor and covariant derivative corresponding to the
two-dimensional transverse metric $h_{ij}$.

Concerning the gauge field strengths, their non-vanishing
components are determined by a single magnetic component, that is,
\begin{eqnarray}
    F_{ij} &=& \epsilon_{ij} B,\\
    F^A_{ij} &=& \epsilon_{ij} B^A,
\end{eqnarray}
where we have introduced the covariantly constant tensor field
$\epsilon_{ij}=-\epsilon_{ji}$, normalized so that
$\epsilon_{ij}\epsilon^{jk}=\delta_i^k$.

\section{Bogomol'nyi Equations}

A significant simplification of this system is obtained if
self-duality conditions are satisfied, i.e. if the system admits a
Bogomol'nyi limit \cite{Bogomolny:1975de}. It is well known
\cite{Linet:1987qu}-\cite{Comtet:1987wi} that in the usual Abelian
Higgs model, self-duality properties take place in curved
space-time if $L(x^i)$ and $K(x^i)$ are constants, say, 1. We will
see now that the present generalized Higgs model has also a
Bogomol'nyi limit if we keep
\begin{equation}
       L(x^i)=1,\quad\quad K(x^i)=1.
       \label{nk}
\end{equation}

Using these conditions, it is easy to verify that $R_{ij}=r_{ij}$
and $R=r$, these leading to that $G_{ij}$ vanishes identically.
Einstein equation (\ref{eins}) then implies the vanishing of
$T_{ij}$, that is,
\begin{equation}
     -\frac{1}{2e_1^2}B^2h_{ij}-\frac{1}{2e_1^2}B^2h_{ij}+{\cal D}_i \bar\phi^r{\cal D}_j \phi_r
     +{\cal D}_j \bar\phi^r{\cal D}_i \phi_r  - h_{ij} {\cal D}_k \bar\phi^r {\cal D}^k \phi_r
     +h_{ij} V=0
     \label{conde}
\end{equation}
The latter equation implies
\begin{equation}
     {\cal D}_i \bar\phi^r{\cal D}_j \phi_r +{\cal D}_j \bar\phi^r{\cal D}_i \phi_r  \propto h_{ij}
     \label{con}
\end{equation}
Now, without loss of generality, we may take the two-dimensional
metric $h_{ij}$ to be conformally flat, i.e.
\begin{equation}
     h_{ij}=-\Omega^2 \delta_{ij}
\end{equation}
Then, we can write condition (\ref{con}) as
\begin{equation}
     {\cal D}_z \bar\phi^r{\cal D}_z \phi_r = 0
\end{equation}
with $z=x+iy$. The easiest way to solve this equation is by
requiring either ${\cal D}_z \phi^a_{\;r} = 0$
 or ${\cal D}_z \bar\phi^r_{\;a} = 0$. Changing back to an arbitrary spatial coordinate system, we thus find
the covariant self-duality condition for the Higgs field
\begin{equation}
     {\cal D}_i \phi_r +i\eta\epsilon_{i}^{\; j} {\cal D}_j \phi_r  =0,
     \label{bogohiggs}
\end{equation}
where $\eta=\pm 1$ corresponds to self-dual or anti-self-dual
solutions. Note however that, in contrast with what happens in the
($N_f=1$) Abelian case \cite{Clement:1995cr}, equations (\ref{nk})
do not imply a priori the self-duality equations
(\ref{bogohiggs}).

If we now return to eq.(\ref{conde}) and use the Higgs
self-duality equation, we arrive to the following condition for
the Higgs potential
\begin{equation}
       V = \frac{B^2}{2 e_1^2} + \frac{B^A B^A}{2 e_2^2}.
       \label{pot1}
\end{equation}

In order to get a first order equation for the gauge fields, we
consider the Higgs equations (\ref{higgs}) and notice that for
self-dual Higgs configuration they become
\begin{equation}
      \left(B^A(T^A)_{\;b}^a + \frac{i}{\sqrt{2N_c}} B \delta^a_b\right)\phi^b_{\;r} = i\eta
      \frac{\partial V}{\partial \bar\phi^r_{\; a}}
      \label{pot2}
\end{equation}
Now, thinking of $B$ and $B^A$ as functions of $\phi$, it follows
from the Yang-Mills equations (\ref{yang}),(\ref{gany}) that they
should be cuadratic in $\phi$. In order to satisfy relation
(\ref{pot2}), we will consider a quartic Higgs potential for the
theory. A generic quartic potential which respect both gauge and
flavor invariances can be written as:
\begin{equation}
       V(\phi,\bar\phi) = c_1 + c_2 \bar\phi^r\phi_r + c_3(\bar\phi^r\phi_r)^2 + c_4 (\bar\phi^rT^A\phi_r)^2
\end{equation}
From eq.(\ref{pot2}) one obtains that
\begin{equation}
      B=  \eta\sqrt{2N_c} (c_2 + 2c_3\bar\phi^r\phi_r ), \quad\quad\quad
      B^A = 2i \eta c_4 \bar\phi^rT^A\phi_r.
      \label{vogo}
\end{equation}
Constants $c_\alpha$ ($\alpha=1,...,4$) are determined by
requiring, firstly, that configurations satisfying self-dual
conditions (\ref{bogohiggs}) and (\ref{vogo}) be solutions of the
Yang-Mills equations (\ref{yang}),(\ref{gany}) and secondly, that
the Higgs potential has the minimum for $\bar\phi^r\phi_r=N_c\xi$.
Then, self-duality equations for the gauge fields take the form
\begin{eqnarray}
      B &=&  \eta \frac {e_1^2}{\sqrt {2N_c}} (\bar\phi^r\phi_r - N_c\xi),\nonumber\\
      B^A &=& -i \eta e_2^2 \bar\phi^rT^A\phi_r,
      \label{bogogauge}
\end{eqnarray}
and the potential takes the form
\begin{eqnarray}
      V(\phi,\bar\phi) &=& \frac{e_1^2}{4N_c} ( \bar\phi^r\phi_r - N_c\xi)^2 - \frac{e_2^2}{2}(\bar\phi^rT^A\phi_r)^2.
      \label{V}
\end{eqnarray}
In summary, the self-duality first order equations for the Higgs
and gauge fields are
\begin{eqnarray}
     &&{\cal D}_i \phi_r +i\eta\epsilon_{i}^{\; j} {\cal D}_j \phi_r
     =0,\nonumber\\
     && B = \frac 1 2 \epsilon^{ji} F_{ij} =  \eta \frac {e_1^2}{\sqrt {2N_c}} (\bar\phi^r\phi_r - N_c\xi),\nonumber\\
     && B^A = \frac 1 2 \epsilon^{ji} F^A_{ij}= -i \eta e_2^2 \bar\phi^rT^A\phi_r,
     \label{bogo}
\end{eqnarray}

It is worth commenting that the existence of the first order
equations (\ref{bogo}) is strongly related to the possibility of
having a locally supersymmetric theory whose bosonic sector
coincides with our model. In fact, this supergravity theory could
be used to obtain not only the Bogomol'nyi equations for the
matter fields (\ref{bogo}) but also Bogomol'nyi first-order
equations for the gravitational field \cite{Gibbons:1982fy}. More
precisely, one expects that the Bogomol'nyi equations could be
obtained from the vanishing of the supersymmetric variation of the
fermionic fields. In particular, the supersymmetric transformation
of the gravitino should yields to a first-order Killing spinor
equation for the supersymmetry parameter. Einstein equations will
then be automatically satisfied as a consequence of the
integrability condition of this Killing equation.

Clearly, the Higgs potential (\ref{V}) is positive definite.
Requiring its vanishing leads, due to the first term, to $\phi$ to
develop a vacuum expectation value (VEV) and, due to the second
term, to the VEV to satisfy
\begin{eqnarray}
   \phi^a_{\;r}\bar\phi^r_{\;b} = \xi \delta^a_b.
   \label{condichi}
\end{eqnarray}
Let us discuss shortly how the vacua of the Higgs potential
(\ref{V}) is and its dependence of $N_c$ and $N_f$. It is clear
from (\ref{condichi}) that there is no vacua with vanishing
potential for $N_f<N_c$, so this case is trivial. In the case of
$N_f=N_c$ there is an unique, isolated vacuum which, up to gauge
transformations, takes the form
\begin{eqnarray}
   \phi^a_{\;r}= \sqrt\xi \delta^a_r.
   \label{vaccu}
\end{eqnarray}
The vacuum field (\ref{vaccu}) has the pattern of symmetry
breaking \cite{Hanany:2003hp}-\cite{Auzzi:2003fs}
\begin{equation}
    U(1)\times SU(N_c)\times SU(N_f) \longrightarrow SU(N)_{\rm c+f},
\end{equation}
where the surviving unbroken group $SU(N)_{\rm c+f}$ is a
simultaneous gauge and flavor rotation. Due to this, the theory is
said to lie in the colour-flavor locked phase. Finally, in the
$N_f>N_c$ case the theory has a Higgs branch of vacua, denoted
${\cal N}_{N_c,N_f}$ \cite{Hanany:2003hp}. For example, for
Abelian theories, which support semi-local strings
\cite{Vachaspati:1991dz}-\cite{Hindmarsh:1991jq}, the vacua is
simply ${\cal N}_{1,N_f}={\bf CP}^{N_f-1}$. In general, the Higgs
branch is the Grassmannian of $N_c$ planes in ${\bf C}^{N_f}$,
\begin{equation}
     {\cal N}_{N_c,N_f} = Gr(N_c,N_f) = \frac{SU(N_f)}{U(1)\times SU(N_c)\times SU(N_f-N_c) }
\end{equation}
This is a symmetric space, and we may choose to work in any of the
vacua without loss of generality. We pick,
\begin{eqnarray}
   &&\phi^a_{\;r}= \sqrt\xi \delta^a_r \quad\quad r=1,...,N_c\nonumber\\
   &&\phi^a_{\;r}= 0 \quad\quad\quad \quad r=N_c+1,...,N_f
   \label{vachio}
\end{eqnarray}
In this vacuum, the pattern of symmetry breaking is
\begin{eqnarray}
    U(1)\times SU(N_c)\times SU(N_f) \longrightarrow S[U(N_c)_{\rm c+f}\times U(N_f-N_c)]
\end{eqnarray}
where $S[\otimes_i U(N_i)]$ means we project out the diagonal,
central $U(1)$ from $\otimes_i U(N_i)$. Since the surviving
unbroken group includes $U(N_c)_{\rm c+f}$, in the $N_f>N_c$ case
the theory also lies in the colour-flavor locked phase.

Returning to the equations of motion, it is clear that with the
Higgs potential eq.(\ref{V}), condition (\ref{pot1}) is
automatically satisfied by self-dual gauge configurations. Thus,
the only second-order equations that remain to solve are the $00$
and $33$ components of Einstein equation, which yields both to the
following expression for the two-dimensional Ricci scalar $r$,
\begin{equation}
     r= 16 \pi G ( h^{ij}{\cal D}_i \bar\phi^r{\cal D}_j \phi_r - 2 V)
     \label{rr}
\end{equation}
Using the identity
\begin{eqnarray}
    h^{ij} {\cal D}_i \bar\phi^r{\cal D}_j \phi_r &=& \frac 1 2  h^{ij}({\cal D}_i \bar\phi^r -i\eta\epsilon_{i}^{\; l}
    {\cal  D}_l \bar\phi^r) ({\cal D}_j \phi_r +i\eta \epsilon_{j}^{\; k}{\cal D}_k \phi_r) \nonumber\\
    &-& i\eta B^A
    \bar\phi^rT^A\phi_r + \eta \frac {1}{\sqrt{2N_c}}B \bar\phi^r\phi_r + \eta \sqrt{\frac{N_c}{2}}
    \epsilon^{ik}  j_{i;k}
    \label{iden}
\end{eqnarray}
and the self-duality equations (\ref{bogo}), we can rewrite the
equation (\ref{rr}) for $r$ as,
\begin{equation}
     r = 8\pi \sqrt{2N_c} G \eta (\xi B + \epsilon^{ik}j_{i;k}),
     \label{r}
\end{equation}
where for self-dual configurations the $U(1)$ current can be
written as
\begin{equation}
     j_i= - \frac{\eta}{\sqrt {2N_c}} \epsilon_{i}^{\; j}(\bar \phi^r \phi_r)_{,j}.
\end{equation}
Now, in the conformal coordinate system $r$ takes the simple form
\begin{equation}
       r= \Omega^{-2}\Delta \log \Omega^2
\end{equation}
where $\Delta$ is the flat-space Laplacian, i.e.,
$\Delta=\delta^{ij}\partial_i\partial_j$. From eq.(\ref{r}) we
then get the following equation for $\Omega^2$:
\begin{equation}
    \Delta (\log\Omega^2)=-8\pi G [\Delta(\bar \phi^r \phi_r)+\sqrt{2N_c}\eta \xi F_{12}]
    \label{log h}
\end{equation}

\subsection*{The Bogomol'nyi bound}

As it is well known, the notion of energy in general relativity is
more subtle than in special relativity. In the present case, since
we are considering static axisymmetric matter configurations which
tends asymptotically to their vacuum values, the two-dimensional
transverse metric will tend asymptotically to that of a flat cone.
Therefore, the deficit angle $\delta$ may be taken as a measure of
the gravitational energy per unit length (see \cite{Comtet:1987wi}
and reference therein). In our case, the deficit angle takes the
form,
\begin{eqnarray}
      \frac{\delta}{8\pi G} &=&  \int d^2x \sqrt g T_0^{\; 0}\nonumber\\
      &=&   \int d^2x \sqrt h \left\{ \frac {1}{2e_1^2} B^2 + \frac {1}{2e_2^2} B^A B^A -
      h^{ij} {\cal D}_i \bar\phi^r{\cal D}_j \phi_r + V \right\}
\end{eqnarray}
Now, by means of identity (\ref{iden}) and the Bogomol'nyi trick,
the energy density $T_0^{\; 0}$ can be re-written as
\begin{eqnarray}
      T_0^{\; 0} &=& \frac {1}{2e_1^2} [B - \eta \frac {e_1^2}{\sqrt {2N_c}} (\bar\phi^r\phi_r - N_c\xi)]^2
      +  \frac {1}{2e_2^2} [B^A + i \eta e_2^2 \bar\phi^rT^A\phi_r]^2 \nonumber\\
      &+& \frac{1}{2\sqrt h}
      |{\cal D}_i \phi_r +i\eta \epsilon_{i}^{\; j}{\cal D}_j \phi_r|^2 - \eta \sqrt{\frac{N_c}{2}} \xi B
      + \eta \sqrt{\frac{N_c}{2}} \epsilon^{ki} j_{i;k}
      \label{t00}
\end{eqnarray}
Thus, integration of (\ref{t00}) yields a Bogomol'nyi bound for
the energy, i.e.
\begin{equation}
       \frac{\delta} {8\pi G}\geq \xi |\Phi|,
       \label{cota}
\end{equation}
where $\Phi$ is a topological number defined by
\begin{equation}
      \Phi = \sqrt{\frac {N_c}{2}}\int d^2x \sqrt h B  = \sqrt{\frac {N_c}{2}}\oint A_i d\sigma^i=2\pi n.
\end{equation}
As expected, we can see that the bound is saturated by
configurations satisfying self-duality equations (\ref{bogo}).

\section{Non-Abelian Local Strings - $N_c=N_f=N$}

In order to find non-Abelian vortex solutions to the self-duality
equations, let us consider, following \cite{Auzzi:2003fs},
rotationally symmetric configurations through the ansatz:
\begin{eqnarray}
     \phi &=& \left(
          \begin{array}{ccccc}
          \varphi(r) & 0 & \cdots & 0 & 0 \\
           0& \varphi(r) & \cdots & 0& 0 \\
          \vdots & \vdots & \ddots & \vdots\\
           0 & 0& \cdots & \varphi(r)& 0 \\
           0& 0 & \cdots & 0& e^{in\theta}\tilde\varphi(r) \\
          \end{array}
          \right),\nonumber\\
     \nonumber\\
     A_i^A T^A &=& \left(
          \begin{array}{ccccc}
          \;\;\,1\;\; & 0& \cdots & 0& 0 \\
           0& \;\;\,1 \;\; & \cdots & 0 & 0\\
          \vdots & \vdots & \ddots &\vdots & \vdots\\
           0 & 0& \cdots & \;\;\,1 \;\;& 0 \\
           0& 0 & \cdots & 0& -(N-1)\\
          \end{array}
          \right)\frac{i}{N}(\partial_i \theta) (f_N(r)-n), \nonumber\\
     \nonumber\\
     A_i &=& \sqrt{\frac{2}{N}} (\partial_i \theta) (-f(r)+n),
     \label{rota}
\end{eqnarray}
where $(r,\theta)$ are the polar coordinates in the
two-dimensional transverse space.

Inserting this ansatz in the self-duality equations (\ref{bogo})
we arrive to the first-order differential equations satisfied by
the profile functions
\begin{eqnarray}
    &&r\partial_r \varphi+\frac {\eta} {N} (f-f_N)\varphi
    =0 \nonumber\\
    &&r\partial_r \tilde\varphi+\frac {\eta} {N} (f-(1-N)f_N)\tilde\varphi
    =0\nonumber\\
    &&\frac 1 r \partial_r f + \eta \frac{e_1^2}{2}\Omega^2 ((N-1)\varphi^2 + \tilde\varphi^2 - N\xi)=0\nonumber\\
    &&\frac 1 r \partial_r f_N + \eta \frac{e_2^2}{2}\Omega^2 (\tilde\varphi^2 -\varphi^2)=0
    \label{prof eq}
\end{eqnarray}
The boundary conditions at the origin follows from the requirement
that the fields be nonsingular. This implies that
\begin{eqnarray}
   n\tilde\varphi(0)=0,\quad f_N(0)=n,\quad f(0)=n.
\end{eqnarray}
At spatial infinity, configurations have to tend asymptotically to
their vacuum values and then
\begin{eqnarray}
   \varphi(\infty)=\tilde\varphi(\infty)=\sqrt
   \xi,\quad f (\infty)=f_N (\infty)=0
\end{eqnarray}
The first and second equation of (\ref{prof eq}) can be solved for
the profiles of the gauge fields
\begin{eqnarray}
       && f = \frac \eta 2 r \partial_r ((1-N)\log \varphi^2 - \log \tilde\varphi^2), \nonumber\\
       && f_N = \frac \eta 2 r \partial_r (\log \varphi^2 - \log \tilde\varphi^2)
       \label{prof}
\end{eqnarray}
We expect $\tilde \varphi$ to have a zero only at $r=0$, whereas
$\varphi$, which does not wind, to have no zeros. Therefore,
eq.(\ref{prof}) will be valid everywhere outside the origin.

Concerning field $\Omega^2$, after using eqs.(\ref{rota}) and
(\ref{prof}) its equation of motion (\ref{log h}) becomes
\begin{equation}
    \Delta \{ \log\Omega^2 +8\pi G [(N-1)\varphi^2+\tilde\varphi^2 - \xi \log(\varphi^{2(N-1)}\tilde\varphi^2)]\}=0
\end{equation}
For a charge-$n$ vortex solution, $\tilde\varphi$ will behave as
$\tilde\varphi \sim  {\rm const} \,r^{|n|}$ if $r \rightarrow 0$.
It then follows that
\begin{equation}
    \log\Omega^2 + 8\pi G [(N-1)\varphi^2+\tilde\varphi^2 - \xi \log(\frac{\varphi^{2(N-1)}\tilde\varphi^2}{r^{2|n|}})]
   \label{raiz}
\end{equation}
is harmonic and bounded, hence is a constant. In particular this
implies that the conformal factor has the following behavior at
infinity
\begin{equation}
    \Omega^2 \sim {\rm const} \;r^{2(B-1)}\quad\quad {\rm if} \;r \rightarrow \infty,
    \label{asymp}
\end{equation}
where $B=1-8\pi |n|G\xi $. Fixing the constant in eq.(\ref{asymp})
to be $\xi^{B-1}$, we can write the asymptotic form of the metric
as
\begin{equation}
       ds^2 \sim dt^2 - (dx^3)^2 - (\xi r^2)^{B-1}(dr^2 + r^2 d\theta^2)
\end{equation}
Thus, we find that the metric corresponding to a single
non-Abelian local string has the same asymptotic behavior as that
of the gravitating Abelian string (given by eq.(\ref{cono})). It
is characterized by the dimensionless parameter $G\xi$, which
determines the strength of the gravitational coupling of the
string \cite{Garfinkle:1985hr}. If $G\xi < \frac {1}{8\pi}$ (i.e.
$B>0$), the metric is asymptotically conical. This can be easily
seen by considering a new radial coordinate $\rho$ given by $\sqrt
\xi \rho = B^{-1} (\sqrt \xi r)^B $, which yields to the
Minkowskian form (\ref{cono}) for the asymptotic metric:
\begin{equation}
       ds^2 \sim dt^2 - (dx^3)^2 - d\rho^2 - B^2\rho^2 d\theta^2.
       \label{minko}
\end{equation}
The deficit angle corresponding to the metric (\ref{minko}) is
\be
    \delta = 2\pi(1-B)=16 \pi^2|n| G\xi.
\ee
As the symmetry breaking scale grows, $\delta$ exceeds $2\pi$ and
the conical picture of the string space-time must be abandoned.
For a critical string (with $G\xi = 1/8\pi$ and $\delta = 2\pi$),
the two-dimensional space is like a cylinder at infinite. Finally,
over-critical strings have a deficit angle greater than $2\pi$,
which happens for $G\xi
> 1/8\pi$. This means that at infinity the space is like
an inverted cone, closing up with a conical singularity which is
at a finite proper distance. Those strings having a deficit angle
$\delta\geq 2\pi$ are known as supermassive strings
\cite{Ortiz:1990tn}. Due to the presence of the singularity in the
metric, supermassive strings appear to be of little physical
interest.

Since the remaining Einstein equation can be integrated
explicitly, we are thus left with a system of coupled equations
for the Higgs profile functions,
\begin{eqnarray}
    &&\Delta \log(\varphi^{2(N-1)}\tilde\varphi^2) = e_1^2 \Omega^2 ((N-1)\varphi^2 + \tilde\varphi^2 - N\xi)
   \vphantom{\frac 1 2}\nonumber\\
    &&\Delta \log(\varphi^{-2}\tilde\varphi^2) =e_2^2 \Omega^2 (\tilde\varphi^2 -\varphi^2),
\end{eqnarray}
where $\Omega^2$ is determined from eq.(\ref{raiz}).
Unfortunately, as in the flat case $G=0$ we have not been able to
find analytical solutions. We can, however, establish from
eq.(\ref{prof eq}) the asymptotic behavior of the solutions near
$r=0$ and for large $r$. Near the polar axis, the first terms of
the development in a power serie in $r$ are
\begin{eqnarray}
       \varphi &\sim& \varphi_0  + \frac{\varphi_0}{8N}\left((e_2^2-e_1^2)\varphi_0^2 + e_1^2N(\varphi_0^2-\xi)\right )r^2,
       \nonumber\\
       \tilde\varphi &\sim&  \tilde\varphi_0 r^{|n|},\vphantom{\frac 1 2}\nonumber\\
       f &\sim & n +\frac \eta 4 e_1^2\Omega^2_0(\varphi_0^2-N(\varphi_0^2-\xi) )r^2,\nonumber\\
       f_N &\sim & n +\frac \eta 4 e_2^2 \Omega^2_0\varphi_0^2r^2,\quad\quad r \rightarrow 0
\end{eqnarray}
where $\varphi_0$ and $\tilde\varphi_0$ are two arbitrary
constants and $\Omega^2_0=\Omega^2(r=0)$ could be determined
through (\ref{raiz}). Concerning the behavior for large $r$, the
profile functions are modified by the non-trivial metric.
Nevertheless, by rescaling the coordinates to be Minkowskian one
can obtain the usual exponential behavior of the ANO strings
\cite{Nielsen:1973cs}. Thus, using the radial coordinate $\rho$
defined through $\sqrt \xi \rho = B^{-1} (\sqrt \xi r)^B$ (for
$B>0$), the behavior at large distances results
\begin{eqnarray}
       \varphi &\sim& \sqrt \xi +\varphi_\infty \rho^{-\frac 1 2}(e^{-M_1\rho}-e^{-M_2\rho}),
       \vphantom{\frac 1 2}\nonumber\\
       \tilde\varphi &\sim& \sqrt \xi +\varphi_\infty \rho^{-\frac 1 2}(e^{-M_1\rho}-(1-N)e^{-M_2\rho}),
       \vphantom{\frac 1 2}\nonumber\\
       f &\sim & \varphi_\infty \eta N e_1 B \rho^{\frac 1 2}e^{-M_1\rho},\vphantom{\frac 1 2}\nonumber\\
       f_N &\sim & \varphi_\infty \eta N e_2 B \rho^{\frac 1 2}e^{-M_2\rho},\quad\quad \rho\rightarrow\infty
       \vphantom{\frac 1 2}
       \label{expo}
\end{eqnarray}
where $\varphi_\infty$ is an arbitrary constant and
$M_i=e_i\sqrt\xi$ ($i=1,2$). The dominant behavior of $\varphi$,
$\tilde\varphi$ is given by the smallest exponential in
(\ref{expo})

Let us now discuss some facts about the vortex moduli space. While
the vacuum is $SU(N)_{\rm c+f}$ symmetric, the solution given by
eq.(\ref{rota}) breaks this symmetry down to $U(1)\times SU(N-1)$.
This means that there exist a set of solutions with the same
topological charge parameterized by the coset
\cite{Hanany:2003hp},\cite{Auzzi:2003fs}
\begin{eqnarray}
   \frac{SU(N)_{\rm c+f}}{SU(N-1)\times U(1)} \cong {\bf CP}^{N-1}
\end{eqnarray}
Thus, if we suppose that the center-of-mass collective coordinates
are decoupled, the moduli space in the case of a single unit
charge vortex takes the form
\begin{eqnarray}
   {\cal M}\cong {\bf C}\times{\bf CP}^{N-1}
\end{eqnarray}
where ${\bf C}$ parameterizes the center of mass of the vortex
configuration. The presence of these extra orientational
collective coordinates makes the vortices genuinely non-Abelian.
One can  make explicit the non-Abelian nature of the solution
(\ref{rota}) by applying the color-flavor rotation preserving the
asymmetric vacuum. To this end, it is convenient first to pass to
the singular gauge where the scalar fields have no winding at
infinite, while the vortex flux comes from the vicinity of the
origin. Then, the Higgs and gauge fields can be written as
\begin{eqnarray}
     \phi &=& {\cal U} \left(
          \begin{array}{ccccc}
           \varphi(r)& 0 & \cdots & 0 & 0 \\
           0& \varphi(r) & \cdots & 0& 0 \\
          \vdots & \vdots & \ddots & \vdots& \vdots \\
           0 & 0& \cdots & \varphi(r)\\
           0& 0 & \cdots & 0& \tilde\varphi(r) \\
          \end{array}
          \right){\cal U}^{-1},\nonumber\\
     \nonumber\\
     A_i^A T^A &=& {\cal U}\left(
          \begin{array}{ccccc}
          \;\;\,1\;\; & 0& \cdots & 0& 0 \\
           0& \;\;\,1 \;\; & \cdots & 0 & 0\\
          \vdots & \vdots & \ddots &\vdots & \vdots\\
           0 & 0& \cdots & \;\;\,1 \;\;& 0 \\
           0& 0 & \cdots & 0& -(N-1)\\
          \end{array}
          \right){\cal U}^{-1}\frac{i}{N}(\partial_i \theta) f_N(r), \nonumber\\
     \nonumber\\
     A_i &=& -\sqrt{\frac{2}{N}} (\partial_i \theta) f(r),
     \label{rota2}
\end{eqnarray}
where ${\cal U}\in SU(N)$ parameterizes the orientational
collective coordinates associated with the flux rotation in
$SU(N)$. Following \cite{Gorsky:2004ad}, we can parameterize these
matrices as follows:
\begin{eqnarray}
   \frac{1}{N}\left\{{\cal U}\left(
          \begin{array}{ccccc}
          \;\;\,1\;\; & 0& \cdots & 0& 0 \\
           0& \;\;\,1 \;\; & \cdots & 0 & 0\\
          \vdots & \vdots & \ddots &\vdots & \vdots\\
           0 & 0& \cdots & \;\;\,1 \;\;& 0 \\
           0& 0 & \cdots & 0& -(N-1)\\
          \end{array}
          \right){\cal U}^{-1}\right\}^a_{\, b}=-n^a n_b^* + \frac 1 N
          \delta^a_b,
          \label{parame}
\end{eqnarray}
where $n^a$ is a complex vector in the fundamental representation
of $SU(N)$, and
\begin{eqnarray}
   n^*_a n^a = 1\quad\quad a=1,...,N
   \label{const}
\end{eqnarray}
Note that this gives the correct number of degrees of freedom for
the charge-1 vortex case, namely, $2(N-1)$. With this
parameterization the vortex solution (\ref{rota2}) takes the form
\begin{eqnarray}
    \phi^a_{\; b} &=& \frac 1 N
    (\tilde\varphi(r)+(N-1)\varphi(r))\delta^a_b+(\tilde\varphi(r)-\varphi(r))\left(n^a n^*_b - \frac 1
    N\delta^a_b\right)\nonumber\\
    A_i^A(T^A)^a_{\; b} &=&- i  \left(n^a n^*_b - \frac 1
    N \delta^a_b \right)\partial_i \theta f_N(r)\nonumber\\
    A_i &=&-\sqrt{\frac 2 N} \partial_i \theta f(r)
    \label{ansi1}
\end{eqnarray}
Note that the conformal factor $\Omega^2$, as obtained from
eq.(\ref{raiz}), results to be independent of the orientational
collective coordinates $n^a$.

\section{Non-Abelian Semilocal  Strings - $N_c=N$, $N_f=N + N_e$}

We can easily write the extension of the ansatz (\ref{rota}) for
the case $N_f>N_c$ as follows
\begin{eqnarray}
     \phi &=& \left(
          \begin{array}{ccccccccc}
          \varphi(r) & 0 & \cdots & 0 & 0 & \rho^1_{\; 1}(r)&\cdots & \cdots & \rho^1_{\; N_e}(r)\\
           0& \varphi(r) & \cdots & 0& 0 & \rho^2_{\; 1}(r)&\cdots & \cdots& \rho^2_{\; N_e}(r)\\
          \vdots & \vdots & \ddots & \vdots& \vdots &\vdots& \ddots& & \vdots\\
           0 & 0& \cdots & \varphi(r)& 0 & \vdots& & \ddots & \vdots\\
           0& 0 & \cdots & 0& e^{in\theta}\tilde\varphi(r) &\rho^N_{\; 1}(r)& \cdots & \cdots & \rho^N_{\; N_e}(r)\\
          \end{array}
          \right),\nonumber\\
     \nonumber\\
     A_i^AT^A &=& \left(
          \begin{array}{ccccc}
          \;\;\,1\;\; & 0& \cdots & 0& 0 \\
           0& \;\;\,1 \;\; & \cdots & 0 & 0\\
          \vdots & \vdots & \ddots &\vdots & \vdots\\
           0 & 0& \cdots & \;\;\,1 \;\;& 0 \\
           0& 0 & \cdots & 0& -(N-1)\\
          \end{array}
          \right)\frac{i}{N}(\partial_i \theta) (f_N(r)-n), \nonumber\\
     \nonumber\\
     A_i &=& \sqrt{\frac{2}{N}} (\partial_i \theta) (-f(r)+ n),
     \label{semirota}
\end{eqnarray}
With this ansatz, self-duality equations (\ref{bogohiggs}) become
the following first-order equations for the profile functions
\begin{eqnarray}
    &&r\partial_r \varphi+\frac {\eta} {N} (f-f_N)\varphi
    =0 \nonumber\\
    &&r\partial_r \tilde\varphi+\frac {\eta} {N} (f-(1-N)f_N)\tilde\varphi
    =0\nonumber\\
    &&r\partial_r \rho^a_{\; r}+\frac {\eta} {N} (f-f_N)\rho^a_{\; r}=0 \nonumber\\
    &&r\partial_r \rho^N_{\; r}+\frac {\eta} {N} (f-(1-N)f_N- Nn)\rho^N_{\; r}=0
    \label{efes}
\end{eqnarray}
where $a=1,...,N-1$ and $r=1,...,N_e$. We also need to specify the
boundary conditions which will determine the solutions in these
equations. Is not difficult to see that in order to have
nonsingular fields which tend asymptotically to vacuum
configurations, the boundary condition for the Higgs profile
functions are
\begin{eqnarray}
    && n\tilde\varphi(0)=0,\quad \varphi(\infty)=\tilde\varphi(\infty)=\sqrt \xi,\nonumber\\
    &&\rho^a_{\; r}(\infty)=\rho^N_{\; r}(\infty)=0.
    \label{leo}
\end{eqnarray}
Equations for $\rho^a_{\; r}$ and $\rho^N_{\; r}$ can be solve in
terms of $\varphi$ and $\tilde\varphi$ through the relations
\begin{eqnarray}
    \rho^a_{\; r}(r) = \chi^a_{\; r}\varphi(r), \quad\quad \rho^N_{\; r}(r) = \chi_{r}\frac{\tilde\varphi(r)}{r^{|n|}}
    \label{escu}
\end{eqnarray}
where $\chi^a_{\; r}$ and $\chi_{r}$ ($a=1,...,N-1$;
$r=1,...,N_e$) are complex parameters. Now, the first relation in
eq.({\ref{escu}}) can only be compatible with boundary conditions
(\ref{leo}) if $\chi^a_{\; r}$ , and then $\rho^a_{\; r}$, are
identically zero. Concerning gauge fields, the equations for their
profile functions take now the form
\begin{eqnarray}
    &&\frac 1 r \partial_r f + \eta \frac{e_1^2}{2}\Omega^2 ((N-1)\varphi^2 + (1+\frac{\bar\chi^r\chi_r}{r^{2|n|}})\tilde
    \varphi^2
    -N\xi)=0\nonumber\\
    &&\frac 1 r \partial_r f_N + \eta \frac{e_2^2}{2}\Omega^2 ((1+\frac{\bar\chi^r\chi_r}{r^{2|n|}})\tilde\varphi^2
    -\varphi^2)  =0,
    \label{profal eq}
\end{eqnarray}
which should be complemented with the boundary conditions
\begin{eqnarray}
   f_N(0)= f(0)=n,\quad f (\infty)=f_N (\infty)=0
\end{eqnarray}
Therefore, we get for the Higgs and gauge fields a family of
solutions labelled by $N_e$ complex parameters $\chi_r$, which, as
we shall see, determine the size and orientation of the solutions.
These string configurations are not conventional ANO strings, but,
rather, semilocal strings (for a review of their properties and
their relationship to electroweak strings, see
\cite{Achucarro:1999it}). These may be regarded as a hybrid of an
ANO string and a sigma-model lump. As it is clear from
eqs.(\ref{semirota}), (\ref{efes}) and (\ref{profal eq}), when the
$\chi_r$ parameters tend to zero, we reobtain the non-Abelian
local string of the previous section. On the other hand, when
$|\chi_r|$ tends to infinite, solution (\ref{semirota}) becomes a
sigma-model lump on the target space ${\cal N}_{N,N+N_e}$ (see
below). Recall that, while the vortices are supported by
$\Pi_1(U(N))={\bf Z}$, the lumps are supported by $\Pi_2({\cal
N}_{N,N+N_e})={\bf Z}$.

Concerning the space-time metric corresponding to these
configurations, after getting the gauge profile functions $f_N$
and $f$ from the first and second equations of (\ref{efes}), the
equation (\ref{log h}) for the conformal factor $\Omega^2$ can be
written as
\begin{equation}
    \Delta \{ \log\Omega^2 +8\pi G [(N-1)\varphi^2+(1+\frac{\bar\chi^r\chi_r}{r^{2|n|}})
    \tilde\varphi^2 - \xi \log(\varphi^{2(N-1)}\tilde\varphi^2)]\}=0
\end{equation}
Following the same reasoning as in the previous section we can
infer that
\begin{equation}
    \log\Omega^2 + 8\pi G [(N-1)\varphi^2+(1+\frac{\bar\chi^r\chi_r}{r^{2|n|}})\tilde\varphi^2
    - \xi \log(\frac{\varphi^{2(N-1)}\tilde\varphi^2}{r^{2|n|}})]
   \label{semi raiz}
\end{equation}
is a constant. This leads to the same asymptotic behavior at
infinity as that of the local strings, i.e.
\begin{equation}
    \Omega^2 \sim {\rm const} \;r^{2(B-1)}\quad\quad {\rm if} \;r \rightarrow \infty,
    \label{asinto2}
\end{equation}
with $B=1 - 8\pi |n|G \xi$. Therefore, the analysis of the
asymptotic behavior of the string space as a function of $G\xi$
done for local strings (see discussion after eq.(\ref{asymp})) is
also valid to semilocal ones.

As in the case of the local strings, we can extract the asymptotic
behavior of the gauge and Higgs fields from the first order
eqs.(\ref{efes}) and (\ref{profal eq}). Near the polar axis, the
behavior of the profile functions is
\begin{eqnarray}
       \varphi &\sim& \varphi_0  + \frac{\varphi_0}{8N}\left((e_2^2-e_1^2)(\varphi_0^2 - \bar\chi^r\chi_r\tilde\varphi_0^2)
       + e_1^2 N (\varphi_0^2-\xi)\right )r^2,
       \nonumber\\
       \tilde\varphi &\sim&  \tilde\varphi_0 r^{|n|},\vphantom{\frac 1 2}\nonumber\\
       f &\sim & n +\frac \eta 4 e_1^2\Omega^2_0(\varphi_0^2 - \bar\chi^r\chi_r\tilde\varphi_0^2 -N(\varphi_0^2-\xi))r^2,
       \nonumber\\
       f_N &\sim & n +\frac \eta 4 e_2^2 \Omega^2_0(\varphi_0^2- \bar\chi^r\chi_r\tilde\varphi_0^2)r^2,\quad\quad r
       \rightarrow 0
\end{eqnarray}
where $\varphi_0$ and $\tilde\varphi_0$ are arbitrary constants.
In order to study the behavior at large distance, it is convenient
to change coordinates in the same way as was done in the local
string case. Thus, setting $\sqrt \xi\rho=B^{-1}(\sqrt \xi r)^B$
(for $B>0$), the metric takes the asymptotic Minkowskian form
(\ref{minko}) and the behavior of the profile functions at large
$\rho$ result
\begin{eqnarray}
       \varphi &\sim& \sqrt \xi \left( 1 + \frac{2 n^2(e_1^2-e_2^2)}{N e_1^2e_2^2} \xi^{|n|} \bar\chi^r\chi_r
       (B\sqrt \xi\rho)^{-\alpha-2}\right),
       \nonumber\\
       \tilde\varphi &\sim&  \sqrt \xi \left(1 -\frac{1}{2} \xi^{|n|}\bar\chi^r\chi_r(B\sqrt \xi\rho)^{-\alpha}-
       \frac{2 n^2((N-1)e_1^2+e_2^2)} {N e_1^2e_2^2}  \xi^{|n|} \bar\chi^r\chi_r(B\sqrt \xi\rho)^{-\alpha-2}\right),
       \nonumber\\
       f &\sim & n \xi^{|n|}\bar\chi^r\chi_r(B\sqrt \xi\rho)^{-\alpha}- \frac{2\eta (\alpha +2)Bn^2}{e_1^2}
        \xi^{|n|}\bar\chi^r\chi_r (B\sqrt \xi\rho)^{-\alpha-2},
       \nonumber\\
       f_N &\sim & n \xi^{|n|} \bar\chi^r\chi_r(B\sqrt \xi\rho)^{-\alpha}- \frac{2\eta (\alpha +2)Bn^2}{e_2^2}
        \xi^{|n|}\bar\chi^r\chi_r (B\sqrt \xi\rho)^{-\alpha-2},\quad\quad \rho \rightarrow \infty
        \label{asinto}
\end{eqnarray}
where $\alpha=2|n|B^{-1}$. The resulting power-law decrease in the
magnetic field at infinite is a significant departure from the
usual exponential decay of the ANO string, which is associated
with the confinement of magnetic flux \cite{Hindmarsh:1991jq}.
Furthermore, semi-local strings develop additional collective
coordinates $\chi_r$ related to unlimited variations of their
transverse size. The width of the flux tube results then
completely undetermined, instead of being the Compton wave-length
of the vector particle as in the ANO string case. This leads to a
dramatic effect - semi-local strings, in contradistinction to the
ANO strings, do not support linear confinement (see
\cite{Shifman:2006kd} for a nice analysis on deconfinement in the
non Abelian semi-local string context).

In order to parameterize the semi-local string solution
(\ref{semirota}) in terms of the orientational collective
coordinates, we apply a color-flavor $SU(N)_{c+f}$ rotation
preserving the vacuum (\ref{vachio}). After going to the singular
gauge and applying the color-flavor rotation, the gauge and Higgs
fields can be expressed as
\begin{eqnarray}
    \phi^a_{\; r} &=& (\tilde\varphi(r)-\varphi(r))\left(n^a n_r^* - \frac 1
    N\delta^a_r\right)\nonumber\\
    &+& \frac 1 N (\tilde\varphi(r)+(N-1)\varphi(r))\delta^a_r \quad\quad\quad\quad r=1,...,N\nonumber\\
    \phi^a_{\; r} &=& \tilde\varphi(r) \frac{e^{-in\theta}}{r^{|n|}} e^{i\delta} n^a\chi_{r}
    \quad\quad\quad\quad r=N+1,...,N+N_e\nonumber\\
    A_i^A(T^A)^a_{\; b} &=&- i  \left(n^a n_b^* - \frac 1 N \delta^a_b \right)\partial_i \theta f_N(r)\nonumber\\
    A_i &=&-\sqrt{\frac 2 N} \partial_i \theta f(r)
    \label{corr}
\end{eqnarray}
where we have used the same parametrization for the $SU(N)_{c+f}$
matrices as in the previous section. Thus, $n^a$ is a complex
vector in the fundamental representation of $SU(N)$ satisfying
\begin{eqnarray}
   n^*_a n^a = 1\quad\quad a=1,...,N
   \label{constit}
\end{eqnarray}
Besides, the phase $\delta$ is an arbitrary function of the
orientational moduli, i.e. $\delta=\delta(n,n^*)$. This
arbitrariness in the parametrization (\ref{parame}) will be useful
when we study the low-energy effective action.

We can see that, in the case of charge-1 vortex configuration,
eq.(\ref{corr}) gives the solution parameterized in terms of all
the expected degrees of freedom, namely, $2(N-1)$ orientational
collective coordinates given by $n^a$ and $2N_e$ transverse size
collective coordinates given by $\chi_r$ (of course, we have not
considered the two collective coordinates corresponding to the
position of the center of mass).

\subsection*{Grassmannian sigma-model lumps}

As remarked before, in the limit of a very large transverse size
of the string, solution (\ref{semirota}) approximates a
two-dimensional sigma-model instanton on the Higgs branch of vacua
${\cal N}_{N,N+N_e}= Gr(N,N+N_e)$ lifted to four dimensions, i.e.,
a Grassmannian lump.  Is the purpose of this section to get a
deeper insight on this question. Indeed, we will be able, in the
large transverse size limit, of solving in an explicit analytic
form eqs.(\ref{efes}), (\ref{profal eq}) and (\ref{semi raiz}) for
the matter fields and the space-time metric, this yielding to a
direct proof of the previous relation.

In order to do this, let us first assume the equality of the
constant couplings, $e_1 = e_2=e$. This greatly simplifies the
problem without leading to a substantial loss of generality. After
this assumption, it is convenient to define a new profile function
$k(r)= f(r) - f_N(r)$, which in addition to $\varphi$ satisfy the
following equations
\begin{eqnarray}
    &&r\partial_r \varphi+\frac {\eta} {N} k\varphi
    =0 \nonumber\\
    &&\frac 1 r \partial_r k + \eta \frac{e^2}{2}N\Omega^2 (\varphi^2  - \xi)=0
\end{eqnarray}
together with the boundary conditions
\begin{eqnarray}
   k(0)=0,\quad k(\infty)=0,\quad \varphi(\infty)=\sqrt \xi.
\end{eqnarray}
Clearly, the solutions for $k$ and $\varphi$ are those of the
vacuum, that is,
\begin{eqnarray}
    k(r) \equiv 0, \quad\quad \varphi(r)\equiv \sqrt \xi.
    \label{sim}
\end{eqnarray}

Concerning the rest of the equations, after using (\ref{sim}) they
reduce to
\begin{eqnarray}
    && r\partial_r \tilde\varphi+ \eta  f\tilde\varphi
    =0 \nonumber\\
    &&\frac 1 r \partial_r f + \eta \frac{e^2}{2}\Omega^2 ( (1+\frac{\bar\chi^r\chi_r}{r^{2|n|}})\tilde\varphi^2
    -\xi)=0\nonumber\\
    && \log\Omega^2 + 8\pi G [(1+\frac{\bar\chi^r\chi_r}{r^{2|n|}})\tilde\varphi^2
     - \xi \log(\frac{\tilde\varphi^2}{r^{2|n|}})] = {\rm const}
   \label{masa}
\end{eqnarray}
with the boundary conditions
\begin{eqnarray}
   f(0)=n,\quad f(\infty)=0,\quad \tilde\varphi(0)=0,\quad \tilde\varphi(\infty)=\sqrt \xi.
\end{eqnarray}
Note that in the large lump limit, i.e., $\bar \chi^r\chi_r >>
(e^2 \xi)^{-|n|}$, we can take
\begin{eqnarray}
      \frac{\bar\chi^r\chi_r r^{2(|n|-1)}}{e^2 \xi(r^{2|n|}+\bar\chi^r\chi_r)^2}\cong 0.
\end{eqnarray}
Then, the solutions of (\ref{masa}) has the form
\begin{eqnarray}
      \tilde\varphi&=& \sqrt \xi \frac {r^{|n|}}{\sqrt {r^{2|n|}+\bar\chi^r\chi_r}},\label{varva}\\
      f &=& \frac {n\bar\chi^r\chi_r}{ r^{2|n|}+\bar\chi^r\chi_r},\\
      \Omega^2 &=& {\rm const} (r^{2|n|}+\bar\chi^r\chi_r)^{-8\pi G \xi}\vphantom{\frac 1 2}
      \label{varva2}
\end{eqnarray}

We then see that, as long as these expressions are valid, the
Higgs field $\phi$ in eq.(\ref{semirota}) takes the form
\begin{eqnarray}
     \phi = \left(
          \begin{array}{ccccccccc}
          \sqrt \xi & 0 & \cdots & 0 & 0 & 0&\cdots & \cdots & 0\\
           0& \sqrt \xi & \cdots & 0& 0 & 0&\cdots & \cdots& 0\\
          \vdots & \vdots & \ddots & \vdots& \vdots &\vdots& \ddots& & \vdots\\
           0 & 0& \cdots & \sqrt \xi& 0 & \vdots& & \ddots & \vdots\\
           0& 0 & \cdots & 0& e^{in\theta}\tilde\varphi(r) &\chi_{1}\frac{\tilde\varphi(r)}{r^{|n|}}& \cdots & \cdots &
          \chi_{N_e}\frac{\tilde\varphi(r)}{r^{|n|}}\\
          \end{array}
          \right),
\end{eqnarray}
with $\tilde\varphi$ given by eq.(\ref{varva}). Using this
expression for $\phi$ it is easy to verify that the vacuum
condition
\begin{eqnarray}
   \phi^a_{\;r}\bar\phi^r_{\;b} = \xi \delta^a_b.
\end{eqnarray}
is satisfied at any $r$. Thus, the Higgs field define a map from
the plane ${\bf R}^2$ into the vacuum manifold $Gr(N,N+N_e)$. This
map is analytic and of degree $n$, i.e. a charge-$n$ Grassmannian
lump.

\section{Effective action for low-energy string dynamics}

Knowledge of the moduli space of the vortex configurations is a
necessary ingredient in their applications in cosmology, as cosmic
strings. Low-energy dynamics of vortices can be described by using
the geodesic approximation due to Manton \cite{Manton:1981mp}. The
main idea in \cite{Manton:1981mp} is to approximate the classical
dynamics of solitons by their geodesic motion in the space of
static/stationary solutions. In the case of vortex-like solutions,
one method to do this is to assume that the collective coordinates
are slow-varying functions of the string worldsheet coordinates
$t,x_3$. Then, reinserting the ansatz in the original action, the
moduli become fields of a (1+1)-dimensional sigma-model on the
string world-sheet. This was the procedure followed in
\cite{Auzzi:2003fs},\cite{Gorsky:2004ad} to study non-Abelian
vortex dynamics in flat space-time. A generalization of the
Manton's method to the case of gravitating solitons is, at the
moment, not totally well developed (see recent works on a formal
treatment in \cite{de Wit:2002xz} and some previous applications
to black-holes and ${\bf CP}^{1}$ lumps in
\cite{Gibbons:1986cp},\cite{Speight:1999ad}).

\subsection*{World-sheet theory for local strings}

In the case of local vortices ($N_f=N_c$), we consider  ansatz
eq.(\ref{ansi1}) assuming that the orientational moduli are slowly
varying functions of the string world-sheet coordinates, i.e.
$n^a=n^a(t,x^3)$. Substituting the proposal (\ref{ansi1}) in the
action (\ref{action}) and performing the integral over the $(x^1,
x^2)$ plane, we end with a two-dimensional sigma-model for the $n^a$
fields. Since $n^a$ parameterize the string zero modes, no potential
term is expected to be present in this sigma-model.

Now,  since moduli parameters enter in (\ref{ansi1}) through a
color-flavor rotation  which now gets a dependence on the $t$ and
$x^3$ coordinates, the original ansatz should be complemented by
adding non-trivial   $A_0$ and $A_3$  components to the gauge
potential. Following \cite{Gorsky:2004ad} we propose
\be
   A_\alpha^A (T^A)^a_{\; b}= (\partial_{\alpha} n^a n^*_b -
   n^a\partial_{\alpha} n^*_b - 2n^a n^*_b (n^*\partial_{\alpha}
   n))\rho(r)\quad\quad\quad\quad \alpha=0,3
   \label{compo}
\ee
where   a new profile function $\rho(r)$ has been introduced, to be
  determined by its  equation of motion through a
minimization procedure.

From the $SU(N)$ gauge field strength
\be
   F^A_{\alpha i} (T^A)^a_{\; b}= -(\partial_{\alpha} n^a n^*_b -
   n^a\partial_{\alpha} n^*_b - 2n^a n^*_b (n^*\partial_{\alpha}
   n))\partial_i \rho(r)-i(\partial_{\alpha} n^a n^*_b +
   n^a\partial_{\alpha} n^*_b) \partial_ i \theta f_N (1-\rho)
\ee
we see that  $\rho(r)$    has to satisfy
\be
    \rho(0)=1\quad\quad\quad\quad \rho(\infty)=0
    \label{bece}
\ee
in order to have a finite contribution  in
the action.

After inserting the modified ansatz in the action (\ref{action}), we
get the low-energy effective action
\be
   {\cal S}_{eff} = 2\beta \int dt dx^3 (\partial^\alpha n^* \partial_\alpha n -
   (n^* \partial^\alpha n)(\partial_\alpha n^* n))
   \label{cpene}
\ee
The constant coupling $\beta$ is related to the four-dimensional
coupling $e^2_2$ through the relation
\be
   \beta = \frac{2 \pi}{e^2_2} I
\ee
where $I$ is the integral
\be
   I = \int_0^\infty rdr  [(\partial_r \rho)^2 + \frac{1}{r^2} f_N^2 (1-\rho)^2 +
   \frac {e^2_2}{2} \Omega^2 (2(\tilde \varphi - \varphi)^2 (1-\rho) + (\tilde \varphi^2 +
   \varphi^2)\rho^2)]
   \label{I}
\ee

The integral $I$ can be viewed as an action for the profile
function $\rho$. Thus, $I$ is extremized by configurations $\rho$
satisfying the second-order equation
\be
   -\frac{d^2 \rho}{dr^2} - \frac 1 r \frac{d \rho}{dr} - \frac
   {1}{r^2} f_N^2 (1-\rho) + \frac {e_2^2}{2}\Omega^2 ((\tilde\varphi^2 + \varphi^2)\rho
   -(\tilde\varphi- \varphi)^2 ) =0
   \label{ecuro}
\ee
As done in \cite{Shifman:2004dr} in flat space-time, using the
first-order equations (\ref{prof eq}) one can show that the
solution of (\ref{ecuro}) is given by
\be
   \rho = 1 - \frac{\tilde\varphi}{\varphi}
\ee
Besides, this solution satisfies the boundary conditions
(\ref{bece}). Substituting this solution back into the expression
(\ref{I}) for the integral $I$, one can check that this integral
reduces to a total derivative and is given by the flux of the
string. That is,
\bea
   I = \int_0^\infty dr  \left[ 2 \partial_r\left(\frac{\tilde \varphi}{\varphi}\right)
   \left( -\eta \frac{\tilde \varphi}{\varphi}f_N \right) + \left(1-\left(\frac{\tilde
   \varphi}{\varphi}\right)^2\right) \eta \partial_r f_N\right] =
   -\eta \left.\left[ \left(\frac{\tilde \varphi}{\varphi}\right)^2 f_N - f_N
   \right]\right|_0^\infty = |n|
\eea

Recalling the world-sheet effective theory  for the orientational
coordinates given by eq.(\ref{cpene}), it corresponds to the two
dimensional ${\bf CP}^{N-1}$ sigma-model, as was already
anticipated by using symmetry arguments in section IV. This can be
easily seen from the invariance of the action (\ref{cpene}) under
the $U(1)$ gauge transformations
\be
   n^a \rightarrow e^{i \vartheta(t,x^3)} n^a,\quad\quad\quad n^*_a
   \rightarrow e^{-i \vartheta(t,x^3)} n^*_a
\ee
and the constraint $n^a n^*_a=1$ for the fields. As shown in
\cite{Auzzi:2003fs}, the ${\bf CP}^{N-1}$ sigma-model is also the
theory governing the effective vortex dynamics in flat space-time.
As a result, at this level of approximation, the dynamics of the
orientational moduli of a single local string does not seem to be
affected by the presence of the gravitational coupling. Several
properties of the dynamics of the theory depend on how the moduli
space is (like, for instance, the probability of reconnection of
cosmic strings in the low-energy regimen). Thus, an unchanged
moduli space in the case of an arbitrary number of solitons would
imply that there is no variation in the low-energy physics of
gravitating local strings with respect to that of local strings in
flat space-time. We shall see below that this situation changes
considerably in the case of semilocal vortices.

\subsection*{World-sheet theory for semilocal strings}

Let us study now the case $N_f>N_c$. In this case, we have to
consider both the orientational $n^a$ and the size moduli $\chi_r$
as slowly varying functions of the string world-sheet coordinates
$t, x^3$. For simplicity we will take the gauge constant couplings
to be equal, $e_1 =e_2=e$. Then, we can put $f_N=f$ and
$\varphi=\sqrt \xi$ in the expressions of eq.(\ref{corr}) for the
fields in the singular gauge. Besides, an ansatz for the
components $A_\alpha$ ($\alpha=0,3$) of the gauge field must be
proposed, so we will consider the same expression (\ref{compo}) as
in the local string case. Concerning the space-time metric, due to
the dependence of the conformal factor $\Omega^2$ with the size
moduli $\chi_r$ (see eq.(\ref{semi raiz})), in the case of
semilocal strings also the metric become world-sheet coordinate
dependent.

Thus, introducing the ansatz (\ref{corr}) and (\ref{compo}) in the
action (\ref{action}) we arrive to the effective action for the
moduli coordinates. The corresponding Lagrangian ${\cal L}_{eff}$
can be decomposed in two parts,
\be
    {\cal L}_{eff} =  {\cal L}_{\chi,n}+ {\cal L}_{\chi},
\ee
the first part ${\cal L}_{\chi, n}$ given the effective action for
the orientational coordinates $n^a$ and the second one ${\cal
L}_{\chi}$ including the kinetic terms of the size moduli $\chi_r$
and being independent of $n^a$. The expressions for these
Lagrangians are
\bea
    {\cal L}_{\chi, n} &=&  2 \beta I_0 (\partial^\alpha n^*
     \partial_\alpha n
    -(n^* \partial^\alpha n)(\partial_\alpha n^* n))\vphantom{\left( \frac 1 r \right)^2}\nonumber\\
    &+& \beta I_1 (n^* \partial^\alpha n + i\partial^\alpha
    \delta) (\partial_\alpha \bar\chi^r \chi_r -  \bar\chi^r
    \partial_\alpha\chi_r +  \bar\chi^r\chi_r(n^* \partial_\alpha n + i\partial_\alpha
    \delta))\vphantom{\left( \frac 1 r \right)^2}
    \label{lagas2}\\
    {\cal L}_{\chi} &=&-\frac{1}{8G}\int_0^\infty rdr
    \left( \frac 1 2 \Omega^{-2}\partial^\alpha (\Omega^2) \partial_\alpha
    (\Omega^2) -2\partial^{\alpha}\partial_{\alpha}(\Omega^2) + \frac 1 r\partial_r(r \partial_r (\log
    \Omega^2))  \right)\nonumber\\
    &+& \beta\int_0^\infty rdr\left\{\vphantom{\left( \frac 1 r \right)^2}\frac
    {1}{r^2} \partial^{\alpha} f \partial_{\alpha} f + e^2\Omega^2 \left [
     \partial^{\alpha} \tilde\varphi \partial_{\alpha} \tilde\varphi + \frac{1}{r^{2|n|}}
    \partial^{\alpha} (\tilde\varphi \bar\chi^r) \partial_{\alpha}(\tilde\varphi \chi_r) \right]
    \right.\nonumber\\
    &-& \left. e^2 \left [\partial_r \tilde\varphi \partial_r \tilde\varphi + \left( \frac 1 r f \tilde
    \varphi\right)^2  +(\partial_r \tilde\varphi - |n| \frac 1 r \tilde \varphi)^2\frac{\bar\chi^r\chi_r}{r^{2|n|}}
    + \left( \frac 1 r (f-n) \tilde \varphi\right)^2\frac{\bar\chi^r\chi_r}{r^{2|n|}} \right] \right\} \label{lagas1}
\eea
where $\beta = 2\pi/e^2$ and $I_i$ $(i=0,1)$ are the integrals
given by
\bea
   I_0 &=& \int_0^\infty rdr  [(\partial_r \rho)^2 + \frac{1}{r^2} f^2 (1-\rho)^2 +
   \frac {e^2}{2} \Omega^2 (2(\tilde \varphi - \sqrt \xi)^2 (1-\rho) + (\tilde \varphi^2 +
   \xi)\rho^2 +  \tilde \varphi^2 \frac{\bar\chi^r\chi_r}{r^{2|n|}} (1-\rho)^2]
   \label{I tilde}\\
   I_1 &=& e^2 \int_0^\infty r
    dr \frac {\Omega^2}{r^{2|n|}}\tilde \varphi^2
\eea

The first term in ${\cal L}_{\chi, n}$ coincides with the Lagrangian of a two
dimensional ${\bf CP}^{N-1}$ sigma-model for the fields $n^a$
multiplied by the integral $I_0$ depending on the size moduli
$\chi_r$. On the other hand, the second term in ${\cal L}_{\chi,
n}$ includes mixed kinetic terms between the orientational moduli
$n^a$ and the semilocal size $\chi_r$, which are similar to those
found in \cite{Eto:2007yv}. In our case, the second term in ${\cal
L}_{\chi, n}$ can be eliminated by choosing the phase $\delta$
such that
\be
    \partial_{\alpha} \delta = i n^* \partial_\alpha n.
    \label{delta}
\ee

Concerning ${\cal L}_{\chi}$, the first line in (\ref{lagas1})
represents the contribution of the Hilbert-Einstein term in
(\ref{action}), while the second and third lines come from the
Yang-Mills-Higgs sector. It is clear that from the first line in
${\cal L}_{\chi}$ only the first term has to be considered, since
the rest are total derivatives. Besides, one can put the conformal
factor $\Omega^2$ in terms of $\tilde \varphi$ and $\chi_r$ just
by using the relation (\ref{semi raiz}).

As we did in the $N_f=N_c$ case, we consider the integral $I_0$ as
an action for the profile function $\rho$. Thus, from $I_0$ one
obtains the second-order equation which the function $\rho$ must
satisfy, namely,
\be
   -\frac{d^2 \rho}{dr^2} - \frac 1 r \frac{d \rho}{dr} - \frac
   {1}{r^2} f^2 (1-\rho) + \frac {e^2}{2}\Omega^2 ((\tilde\varphi^2 + \xi)\rho
   -(\tilde\varphi- \sqrt \xi)^2 - \tilde \varphi^2\frac{\bar\chi^r\chi_r}{r^{2|n|}}(1-\rho)) =0
   \label{ecuro}
\ee
with the boundary conditions $\rho(0)=1$, $\rho(\infty)=0$. This
equation is solved by
\be
   \rho = 1 - \frac{\tilde\varphi}{\sqrt \xi}
\ee
as one can show using the first-order equations (\ref{efes}) and
(\ref{profal eq}) for the profile functions. Substituting this
solution back into $I_0$, one finds that this integral reduce to
\bea
   I_0 &=& \int_0^\infty dr  \left[ 2 \partial_r\left(\frac{\tilde \varphi}{\sqrt \xi}\right)
   \left( -\eta \frac{\tilde \varphi}{\sqrt \xi}f \right) + \left(1-\frac{\tilde
   \varphi^2}{ \xi}\right) \eta \partial_r f + r \frac {e^2}{2} \Omega^2
   \frac{\bar\chi^r\chi_r}{r^{2|n|}}\tilde\varphi^2\right] \nonumber\\
   &=& |n| + \frac {e^2}{2}
   \int_0^\infty r dr  \Omega^2
   \frac{\bar\chi^r\chi_r}{r^{2|n|}}\tilde\varphi^2
\eea
Using the latter expression for $I_0$ and the phase $\delta$ given
by eq.(\ref{delta}), the Lagrangian ${\cal L}_{\chi,n}$ takes now
the simpler form
\be
    {\cal L}_{\chi, n} =  2 \beta (|n| + \frac {e^2}{2}
   \int_0^\infty r dr  \Omega^2
   \frac{\bar\chi^r\chi_r}{r^{2|n|}}\tilde\varphi^2) (\partial^\alpha n^*
     \partial_\alpha n
    -(n^* \partial^\alpha n)(\partial_\alpha n^* n))
    \label{lagas3}
\ee

At this point we are ready to face an important question about
semilocal strings, which is the normalizability of the zero modes.
Recently, the moduli space of semilocal non-Abelian strings in
flat space-time ($\Omega^2 = 1$) was obtained through the Manton
procedure in
\cite{Eto:2006db},\cite{Shifman:2006kd},\cite{Eto:2007yv}.
Although some disagreements between the results of these works,
all of them found that some of the orientational and size zero
modes are non-normalizable. In fact, they found that a single
semilocal vortex always has only non-normalizable moduli. The
non-normalizability of a zero mode manifests through an infinite
kinetic term for this mode, due to logarithmic divergences in the
infrared. It follows then that the corresponding moduli become
frozen in this approximation, since any change in it is impeded by
infinite inertia. On the other hand, it was noted  in the case of
self-gravitating ${\bf CP}^{1}$ lumps \cite{Speight:1999ad} that
the deformation of the space-time introduced by gravity is just
sufficient to remove the singularity. One expects, therefore, that
also in the case of semilocal strings the previously frozen moduli
defrost, once gravitational effects are taken into account
\footnote{The author is profoundly indebt with Muneto Nitta for
drawing his attention on the issue of normalizability of zero
modes for gravitating solitons.}. In fact, one can see that this
is what actually happens by using the asymptotic expansions in the
infrared obtained from eqs.(\ref{asinto2}) and (\ref{asinto}),
\bea
    \tilde\varphi &\sim& \sqrt \xi - \frac{\sqrt \xi}{2}\frac {\bar\chi^r \chi_r}{r^{2|n|}}\nonumber\\
    f &\sim& n \frac {\bar\chi^r \chi_r}{r^{2|n|}}\nonumber\\
    \Omega^2 &\sim& {\rm const} \;r^{2(B-1)}\quad\quad\quad\quad {\rm if} \;r \rightarrow
    \infty\vphantom{\frac{\sqrt \xi}{2}}
\eea
where $B=1 - 8\pi|n| G \xi$ (note that these expansions are valid
if $B>0$). Thus, introducing these expansions in the expressions
(\ref{lagas1}) and (\ref{lagas3}) for the effective Lagrangians,
it is easily verified that all modes become normalizable when
strings are coupled to gravity.

Coming back to ${\cal L}_{\chi}$, in order to write this
Lagrangian as a sigma-model one, it would be necessary to put
expression (\ref{lagas1}) explicitly in terms of the field
$\chi_r$. Unfortunately, this is not possible since it is not
known how the profile functions $f$ and $\tilde\varphi$ depend on
the size moduli. We can, however, make use of the large transverse
size limit
\be
    \bar \chi^r\chi_r m_W^2>> 1
\ee
(where we have taken the winding $n$ to be $1$ and called $m_W^2 =
e^2 \xi$). As was shown previously, in this limit we have explicit
analytic solutions given by eqs.(\ref{varva})-(\ref{varva2}), not
only for the matter fields, but also for the space-time metric.
Concerning this last field, we choose the arbitrary constant in
(\ref{varva2}) so that the conformal factor takes the form
\be
   \Omega^2 = \frac{1}{(m_W^2(r^2 + \bar\chi^r\chi_r ))^{8\pi G \xi}}
\ee
It is convenient to recall that the behavior of the corresponding
two-dimensional metric $h_{ij}= \Omega^2 \delta_{ij}$ depends on
the value of the parameter $G\xi$ (or, equivalently, on the value
of the deficit angle $\delta$). If $G\xi < 1/8\pi$, this metric is
asymptotically conical, with a deficit angle $\delta = 16 \pi^2
G\xi$. If $G\xi = 1/8\pi$, the deficit angle is $2\pi$ and then
the two-dimensional transverse space is asymptotically
cylindrical. Finally, a deficit angle greater than $2\pi$, which
happens for $G\xi > 1/8\pi$, means that at infinity the space is
like an inverted cone, with a conical singularity at a finite
proper distance.

Thus, introducing the expressions (\ref{varva})-(\ref{varva2}) for
the profile functions in (\ref{lagas1}) and (\ref{lagas3}), we
obtain the following effective Lagrangians ${\cal L}_{\chi}$ and
${\cal L}_{\chi,n}$:
\bea
   \beta^{-1}{\cal L}_\chi &=&  \left(\frac{1}{12} - \frac{(9-B)(m_W^2 \bar\chi^r\chi_r)^{B}}{8(2-B)}\right)
   \frac{\partial^\alpha(\bar\chi^s\chi_s)\partial_\alpha(\bar\chi^t\chi_t)}
   {(\bar\chi^u\chi_u)^2}+\frac{(m_W^2 \bar\chi^r\chi_r)^{B}}{1-B}\frac{\partial^\alpha\bar\chi^s\partial_\alpha\chi_s}
   {\bar\chi^t\chi_t} - m_W^2\nonumber\\
   \beta^{-1}{\cal L}_{\chi, n} &=&  2  \left( 1 + \frac{(m_W^2 \bar\chi^r\chi_r)^B}{4(1-B)}\right)
   (\partial^\alpha n^* \partial_\alpha n -(n^* \partial^\alpha n)(\partial_\alpha n^* n))
   \label{casi}
\eea
Finally, depending again on the value of the parameter $G\xi$,
three different theories result from (\ref{casi}) for the
low-energy effective dynamics of the vortex in  the large
transverse size limit:

~

$\bullet$ First case: $0< B < 1$ or $G\xi < \frac {1}{8\pi}$,
metric asymptotically conical
\be
   \beta^{-1}{\cal L}_{eff} = \frac{(m_W^2 \bar\chi^r\chi_r)^{B}}{1-B}
   \left( \frac{\partial^\alpha\bar\chi^s\partial_\alpha\chi_s}
   {\bar\chi^t\chi_t} -\frac{(1-B)(9-B)}{8(2-B)} \frac{\partial^\alpha(\bar\chi^s\chi_s)\partial_\alpha(\bar\chi^t\chi_t)}
   {(\bar\chi^u\chi_u)^2} + \frac{1}{2 }(\partial^\alpha n^* \partial_\alpha n
   -(n^* \partial^\alpha n)(\partial_\alpha n^* n))\right)
   \label{1}
\ee

$\bullet$ Second case: $B=0$  or  $G\xi = \frac {1}{8\pi}$, metric
asymptotically cylindrical
\be
    \beta^{-1}{\cal L}_{eff} = \frac{\partial^\alpha\bar\chi^r\partial_\alpha\chi_r}
    {\bar\chi^s\chi_s} - \frac {23}{48}\frac{\partial^\alpha(\bar\chi^r\chi_r)\partial_\alpha(\bar\chi^s\chi_s)}
   {(\bar\chi^t\chi_t)^2}+\frac 5 2 (\partial^\alpha n^* \partial_\alpha n
   -(n^* \partial^\alpha n)(\partial_\alpha n^* n))
   \label{2}
\ee

$\bullet$ Third case: $B<0$ or $G\xi > \frac {1}{8\pi}$, metric
asymptotically conically singular
\be
   \beta^{-1}{\cal L}_{eff} = \frac{1}{12} \frac{\partial^\alpha(\bar\chi^r\chi_r)\partial_\alpha(\bar\chi^s\chi_s)}
   {(\bar\chi^t\chi_t)^2} + 2 (\partial^\alpha n^* \partial_\alpha n
   -(n^* \partial^\alpha n)(\partial_\alpha n^* n))
   \label{3}
\ee

It is interesting to note that for supermassive strings ($G\xi
\geq 1/8\pi$) the dynamics of the orientational modes decouples
from that of the size moduli. Moreover, the Lagrangian for the
orientational moduli of supermassive semilocal strings corresponds
to that of the two-dimensional ${\bf CP}^{N-1}$ sigma-model, as in
the case of local strings.

On the other hand, the most relevant case is the first one, with
$G\xi < 1/8\pi$, since it includes the range of physical
applications (e.g. cosmological data gives an upper limit
$G\xi<10^{-6}$ for cosmic strings -see \cite{1} and references
therein-). In this case, Lagrangian (\ref{1}) presents mixed terms
between size and orientational moduli turning out the theory far
more complex. Indeed, one can expect this kind of terms in the
Lagrangian since our effective theory must be considered as a
deformation of the theory obtained in flat space-time in
\cite{Shifman:2006kd},\cite{Eto:2007yv}.

From expressions (\ref{1})-(\ref{3}) it is clear that in the large
transverse size limit all modes are normalizable, for any value of
the parameter $B$. This could have several consequences in the
low-energy physics of the theory. If a similar defrosting effect
takes place also in the moduli space of more than one string
coupled to gravity, several previous analysis of the dynamics of
strings (like those done in
\cite{Hashimoto:2005hi},\cite{Eto:2006db} where reconnection of
non-Abelian cosmic strings were studied) could be considerably
affected. Thus, we see that, in contrast to what happens for a
single local string, the presence of gravity produce important
changes in the moduli space of semilocal strings, which can be
relevant to the physical properties of this kind of topological
defect.

\section{Summary and Discussion}

The main goal of this work was the construction of a new type of
gravitating string solutions, mainly characterized by being
genuinely non-Abelian. To do this, we considered a four dimensional
Einstein-Yang-Mills-Higgs theory with gauge group $G=U(1)\times
SU(N_c)$, $N_f$ scalar fields in the fundamental representation of
$G$ and an   {\it a priori} undetermined Higgs potential.

Guided by results obtained in the Abelian case
\cite{Linet:1987qu},\cite{Comtet:1987wi}, we proposed an ansatz for
the space-time metric which allows us to find first-order
Bogomol'nyi equations from consistency conditions  resulting from
  the (highly complex) second order Euler-Lagrange equation of
motion.

Not quite surprisingly,   consistency  fixes the Higgs potential
to be the quartic one given by eq.(\ref{V}). In particular, the
resulting potential yields to a pattern of symmetry breaking
containing, in the $N_f \geq N_c$ case, a surviving unbroken group
$SU(N)_{c+f}$.
 This property of the Higgs potential is completely necessary
in order to find strings solutions having an internal
orientational moduli space. Moreover, as it is shown in section
III the gravitational energy (associated to  the total deficit
angle) has, for such a potential, a lower bound which is a
topological number related to the magnetic flux.   {\it A
posteriori}, we verified that the Bogomol'nyi bound is saturated
precisely by configurations satisfying the first order equations.

In order to solve the Bogomol'nyi equations, we proposed a
rotationally symmetric ansatz similar to that used in flat
space-time to get non-Abelian vortices \cite{Auzzi:2003fs}. In the
$N_f=N_c$ case, we showed that this ansatz yields to gravitating
local non-Abelian strings. The non-Abelian character becomes
apparent from the existence of a set of orientational collective
coordinates parameterizing the string solution. Concerning the large
distance behavior, these strings are similar to ANO strings, in the
sense that they have the usual exponential decay.

With respect to the $N_c<N_f$ case, a generalization of the ansatz
allowed us to construct  gravitating semilocal non-Abelian strings.
In this case, semilocal strings acquire, beside the orientational
degrees of freedom, new collective coordinates related to variations
of the transverse size. It is worth noting that this type of strings
has a decreasing power-law large-distances behavior, this resulting
in the deconfinement of the magnetic flux. In fact, the width of the
magnetic flux results completely undetermined since the new
collective coordinates permit to do unlimited variations of the
transverse size of the cosmic string. We were able to find in the
large transverse size limit explicit analytic solutions, not only
for the matter fields, but also for the space-time metric.
Interestingly enough, the explicit solutions allowed us to clearly
show  that semilocal solutions approximate two-dimensional
sigma-model instantons on the Higgs branch of vacua $Gr(N_c,N_f)$ in
the limit of a large transverse size of the string,.

Finally, string world-sheet effective actions for the moduli
coordinates were obtained using the Manton procedure. In the case of
local strings, the dynamics of the orientational moduli turned out
to be that of a two-dimensional ${\bf CP}^{N_c-1}$ sigma-model,
which is just the same effective theory governing the dynamics of
 strings in the flat space-time case.

In contrast,  when semilocal strings are considered,  gravitational
effects already arise at a low-energy level,   radically changing
the moduli dynamics. Indeed, in flat space-time low energy dynamics
of orientational and size moduli is highly constrained due to
non-normalizability of some (and sometimes all) of the zero-modes.
We found that gravity alters completely this situation since, quite
surprisingly, all orientational and size modes previously frozen
become normalizable when strings are coupled to gravity.

In view of the results described above, it would be interesting to
study their relevance concerning the physics of cosmic strings.
For instance, gravity could induce changes in the moduli space
leading to a probability of reconnection $P<1$. To analyze this,
it would be necessary to look for solutions corresponding to
composite gravitating non-Abelian vortices, analogous to those
found in flat space-time in \cite{Auzzi:2005gr}, in order to
obtain the dynamics of two intersecting vortices. In this respect,
it could be quite useful to search also for a generalization on
the moduli matrix approach \cite{Eto:2006pg} to the case of
gravitating solitons. We hope to report on this issues in a
forthcoming work.

\vspace{1 cm}

\noindent\underline{Acknowledgements}: The author wishes to thank
Fidel Schaposnik for constant encouragement and useful discussions
during this work and also to Muneto Nitta and Diego Marques for
very useful comments and discussions on the effective theories.
This work was partially supported by PIP6160-CONICET,
PIC-CNRS/CONICET, BID 1728OC/AR PICT20204-ANPCYT grants and by CIC
and UNLP.

\end{document}